\documentclass[11pt,letterpaper]{article}

\usepackage{amsmath}
\usepackage{amsfonts}
\usepackage{amssymb}
\usepackage{amsthm}
\usepackage[latin1]{inputenc}
\usepackage[spanish, activeacute]{babel}
\usepackage{listings}
\usepackage{hyperref}
\usepackage{soul}
\usepackage[pdftex]{color,graphicx}

\newtheorem{teorema}{Teorema}

\newtheorem{ejemplo}{Ejemplo}
\usepackage{xcolor}
\usepackage[top=3.81cm, bottom=3.81cm, left=2cm, right=1cm]{geometry}

\hypersetup{colorlinks,%
	    citecolor=black,%
	    filecolor=black,%
	    linkcolor=black,%
	    urlcolor=blue}
	    
\title{Una metodología para realizar Diferenciación Automática Anidada}
\author{Juan Luis Valerdi \\ Fernando Raul Rodriguez}
\date{}

\begin{document}
\maketitle

\selectlanguage{spanish} 

\begin{abstract}	

  En este trabajo se presenta una propuesta para realizar
  Diferenciación Automática Anidada utilizando cualquier biblioteca de
  Diferenciación Automática que permita sobrecarga de operadores.
  Para calcular las derivadas anidadas en una misma evaluación de la
  función, la cual se asume que sea anal'itica, se trabaja con el modo
  forward utilizando una nueva estructura llamada SuperAdouble, que
  garantiza que se aplique correctamente la Diferenciación Automática
  y se calculen el valor y la derivada que se requiera.
  \\
  
  This paper proposes a framework to apply Nested Automatic Differentiation
  using any library of Automatic Differentiation which allows operator
  overloading. To compute nested derivatives of a function while it is being 
  evaluated, which is assumed to be analytic, a new structure called SuperAdouble
  is used in the forward mode. This new class guarantees the correct application
  of Automatic Differentiation to calculate the value and derivative of a function
  where is required.
\end{abstract}

\section{Introducción}\label{cha:intro}

 Muchas veces es necesario calcular el valor numérico de una función y
 la vez obtener una aproximación precisa de las derivadas. Ejemplos de
 esto se pueden encontrar en los métodos numéricos de la programación
 no lineal \cite{luenberger}, en los métodos implícitos para la
 resolución numérica de ecuaciones diferenciales \cite{hairer} y en
 problemas inversos en la asimilación de datos \cite{griewank}.

 La Diferenciación Automática (AD, por sus siglas en inglés) es una
 herramienta que puede usarse para calcular las derivadas de cualquier
 función diferenciable, de forma automática. En este contexto,
 ``automática''\ significa que el usuario solo necesita escribir el
 código fuente de la función en un lenguaje determinado, y la
 herramienta puede calcular las derivadas solicitadas sin incurrir en
 errores de truncamiento \cite{griewank}.

 Existen herramientas de Diferenciación Automática para varios lenguajes
 de programación, como ADOL-C \cite{adolc} para programas escritos 
 en C y C++, ADIFOR \cite{adifor} para programas escritos en Fortran, y 
 ADOLNET \cite{adolnet} para lenguajes de la plataforma .NET. 
 
 Una forma de crear estas herramientas de AD en lenguajes que soporten 
 sobrecarga de operadores es crear una clase llamada Adouble que contenga 
 dos campos de tipo double, uno llamado valor y otro derivada. Para esta 
 clase se sobrecargan los operadores aritm'eticos y las funciones elementales, 
 para que, al mismo tiempo que se calculen los resultados de las operaciones, 
 se calculen el valor de las derivadas de esas operaciones \cite{griewank}. 

 Sin embargo, existen ocasiones en las que es necesario calcular unas derivadas
 como paso intermedio para el cálculo de otras y con las t'ecnicas usuales de
 AD esto no es posible.  Un ejemplo sencillo donde se aprecia este fenómeno es 
 el siguiente.

 \begin{ejemplo} \label{ej:anidaad}
  Sea $f$ la función definida como  
  \begin{displaymath}
   f(x)= x^2+\dot{g}(x^3),
  \end{displaymath}
  donde $\dot{g}$ es la función derivada de $g(x)=e^{x^2}.$ Suponiendo que 
  sólo se conoce el código fuente de $g$ y de $f$, se desea calcular el valor y la 
  derivada de $f$ en el punto $x=x_0.$
 \end{ejemplo}

 Estas derivadas que intervienen en el cálculo de otras reciben
 el nombre de derivadas anidadas, y aunque existen herramientas 
 de Diferenciaci'on Autom'atica como ADOL-C \cite{riehme} que 
 permiten el cálculo de estas derivadas, no existe un método 
 general que funcione en todas las herramientas existentes \cite{siskind}, 
 por lo que puede decirse que no existe una solución general para 
 este problema. 
 
 En este trabajo se propone una metodología y una herramienta
 computacional para calcular derivadas anidadas de funciones
 anal'iticas utilizando cualquier biblioteca de AD que soporte
 sobrecarga de operadores.

 La metodolog'ia propuesta en este trabajo parte de la creación de un
 nuevo tipo de dato que se llamar'a SuperAdouble. Al igual que el tipo
 de dato Adouble contendr'a dos campos, uno para almacenar el valor y
 otro para almacenar la derivada. A diferencia de los Adoubles, en el
 que los campos valor y derivada son de tipo double, en los
 SuperAdoubles, estos campos ser'an de tipo Adouble. Para esta nueva
 clase tambi'en se sobrecargan los operadores aritm'eticos y las
 funciones elementales para calcular simult'aneamente el valor de las
 operaciones y sus derivadas.  Al realizar operaciones con estos
 SuperAdoubles es posible calcular derivadas anidadas. 
 
 La fundamentaci'on te'orica de los Adoubles y SuperAdoubles se encuentra
 en \cite{tesis}, donde se realiza una presentación algebraica del espacio de los números
 Adoubles (que son el fundamento de la Diferenciación Automática) y de
 los números SuperAdoubles (que son el fundamento de este trabajo). Tambi'en se
 encuentra en \cite{tesis} una demostraci'on de la posibilidad de calcular derivadas anidadas 
 usando los SuperAdoubles.
 
 Este trabajo se divide en dos secciones: la Secci'on \ref{cha:ad} 
 introduce los conocimientos necesarios para entender e
 implementar la Diferenciaci'on Autom'atica y el modo forward; 
 la Secci'on \ref{cha:metodologia} muestra la metodolog'ia que se
 propone para calcular derivadas anidadas, la clase SuperAdouble
 y su implementaci'on.

\section{Diferenciación Automática}\label{cha:ad}

 En este secci'on se realiza una introducci'on a la diferenciaci'on
 autom'atica, en la que se presentar'an sus modos de aplicaci'on, ejemplos y
 v'ias de implementaci'on. Para la implementaci'on se mostrar'an c'odigos
 en el lenguaje de programaci'on C\# con el objetivo de mostrar el uso de la AD 
 en los ejemplos que se presentan.
 
 \subsection{Introducción a la AD}
  La Diferenciación Automática, también conocida como 
  diferenciación algorítmica, es un conjunto de técnicas y herramientas 
  que permiten evaluar numéricamente la derivada de una función definida 
  mediante su código fuente en un lenguaje de programación.  

  La AD está basada en el hecho de que para evaluar una función en un
  lenguaje de programación dado se ejecutan una secuencia de operaciones
  aritméticas (adición, sustracción, multiplicación y división) y llamados a
  funciones elementales (exp, log, sen, cos, etc.). Aplicando la regla de
  la cadena al mismo tiempo que se realizan estas operaciones, se pueden 
  calcular derivadas de cualquier orden tan exactas como la aritmética 
  de la máquina lo permita \cite{griewank}.

  La base de la AD es la descomposición de diferenciales que provee la regla 
  de la cadena. Para una composición de funciones $f(x) = g(h(x))$ se obtiene, 
  a partir de la regla de la cadena,
  
  \begin{displaymath}
	 \frac{df}{dx} = \frac{dg}{dh} \frac{dh}{dx}.
  \end{displaymath}

  Existen dos modos para aplicar la AD, el modo hacia adelante 
  o modo forward y el modo hacia atrás o modo reverse. El modo forward se obtiene 
  al aplicar la regla de la cadena de derecha a izquierda, 
  es decir, primero se calcula $dh / dx$ y después $dg / dh$, mientras que el modo 
  reverse se obtiene cuando se aplica la regla de la cadena de izquierda a derecha \cite{griewank}.  
 
  A continuaci'on se presentan las ideas fundamentales del modo hacia adelante,
  que por ser m'as sencillo e intuitivo, permite explicar con mayor claridad
  el funcionamiento de la diferenciaci'on autom'atica. El lector interesado
  en el modo hacia atr'as puede consultar \cite{griewank}.

  \subsubsection{Modo forward}

  Los siguientes ejemplos ilustran el funcionamiento del modo forward:

  \begin{ejemplo} \label{ej:ad1}
   Se desea calcular el valor y la derivada de 
   \begin{displaymath}
    f(x) = x^2\cdot\cos(x) 	
   \end{displaymath}
   en el punto $x=\pi$.
  \end{ejemplo}
  
  La siguiente tabla contiene las operaciones necesarias para evaluar esta funci'on
  en una computadora.
  
  \begin{displaymath}
	 \begin{tabular}{|ccccc|}
    \hline
    $w_1$ & = & $x$ & = & $\pi$ \\
    \hline
    $w_2$ & = & $w_1^2$ & = & $\pi^2$ \\
    $w_3$ & = & $\cos(w_1)$ & = & $-1$ \\
    $w_4$ & = & $w_2w_3$ & = & $-\pi^2$ \\
    \hline
    $y$ & = & $w_4$ & = & $-\pi^2$ \\
    \hline
   \end{tabular}
  \end{displaymath}

  Para aplicar el modo hacia adelante, se almacena en una nueva variable
  la derivada de cada operaci'on con respecto a la variable
x. Esta derivada se puede calcular al mismo tiempo 
  que se realiza la operaci'on. La siguiente tabla ilustra este procedimiento.
  
  \begin{displaymath}
	 \begin{tabular}{|ccccc|ccccc|}
    \hline
    $w_1$ & = & $x$ & = & $\pi$ & $\dot{w}_1$ & = & $\dot{x}$ & = & $1$ \\
    \hline
    $w_2$ & = & $w_1^2$ & = & $\pi^2$ & $\dot{w}_2$ & = & $2w_1\dot{w}_1$ & = & $2\pi$ \\
    $w_3$ & = & $\cos(w_1)$ & = & $-1$ & $\dot{w}_3$ & = & $\sin(w_1)\dot{w}_1$ & = & $0$\\
    $w_4$ & = & $w_2w_3$ & = & $-\pi^2$ & $\dot{w}_4$ & = & $w_2\dot{w}_3+\dot{w}_2w_3$ & = & $-2\pi$\\
    \hline
    $y$ & = & $w_4$ & = & $-\pi^2$ & $\dot{y}$ & = & $\dot{w}_4$ & = & $-2\pi$ \\
    \hline
   \end{tabular}
  \end{displaymath}

  En la tabla anterior, las variables $\dot{w}_i$ almacenan la 
  derivada con respecto a x de la operaci'on $w_i$. Como 
  se desea calcular la derivada con respecto a la variable $x$, entonces la nueva variable
  $\dot{w}_1=\frac{dw_1}{dx}$ se inicializa con el valor $1$. Al finalizar el c'alculo, en la variable 
  $\dot{w}_4$ se tiene el valor de la derivada de $f(x)$ con respecto a $x$.
  
  El siguiente ejemplo muestra c'omo se puede aplicar el modo hacia adelante para calcular derivadas
  parciales de funciones de m'as de una variable.

  \begin{ejemplo} \label{ej:ad}
   Se desea calcular el valor y el gradiente de 
   \begin{displaymath}
	  f(x_1,x_2) = x_1x_2+\sen(x_1)
   \end{displaymath}
   en $x_1=\pi$ y $x_2=2.$
  \end{ejemplo}
  
  La función $f$ de este ejemplo puede expresarse mediante las siguientes operaciones.
  \begin{displaymath}
	 \begin{tabular}{|ccccc|}
    \hline
    $w_1$ & = & $x_1$ & = & $\pi$ \\
    $w_2$ & = & $x_2$ & = & $2$ \\
    \hline
    $w_3$ & = & $w_1w_2$ & = & $2\pi$ \\
    $w_4$ & = & $\sen(w_1)$ & = & $0$ \\
    $w_5$ & = & $w_3+w_4$ & = & $2\pi$ \\
    \hline
    $y$ & = & $w_5$ & = & $2\pi$ \\
    \hline
   \end{tabular}
  \end{displaymath}
  
  Para calcular el gradiente utilizando el modo forward hay que realizar el mismo 
  procedimiento del ejemplo anterior, pero en este caso dos veces: una por cada
  derivada parcial. A continuaci'on se muestra el c'alculo de la derivada parcial
  $\frac{df}{dx_1}$. 
  
  \begin{displaymath}
	 \begin{tabular}{|ccccc|ccccc|}
    \hline
    $w_1$ & = & $x_1$ & = & $\pi$ & $\dot{w}_1$ & = & $\dot{x}_1$ & = & $1$ \\
    $w_2$ & = & $x_2$ & = & $2$ & $\dot{w}_2$ & = & $\dot{x}_2$ & = & $0$ \\
    \hline
    $w_3$ & = & $w_1w_2$ & = & $2\pi$ & $\dot{w}_3$ & = & $\dot{w}_1w_2+w_1\dot{w}_2$ & = & $2$ \\
    $w_4$ & = & $\sen(w_1)$ & = & $0$ & $\dot{w}_4$ & = & $\cos(w_1)\dot{w}_1$ & = & $-1$ \\
    $w_5$ & = & $w_3+w_4$ & = & $2\pi$ & $\dot{w}_5$ & = & $\dot{w}_3+\dot{w}_4$ & = & $1$ \\
    \hline
    $y$ & = & $w_5$ & = & $2\pi$ & $\dot{y}$ & = & $\dot{w}_5$ & = & $1$ \\
    \hline
   \end{tabular}
  \end{displaymath}

  En este caso, la variable $\dot{w}_2$ se inicializa con valor $0$ porque
  $\frac{\partial x_2}{\partial x_1}=0, w_2=x_2$ y $\dot{w}_2=\frac{\partial w_2}{\partial x_1}$.
  Al finalizar los c'alculos, en la variable $\dot{w}_5$ se tiene el valor
  de la derivada $\frac{\partial f(x)}{\partial x_1}$.

  El procedimiento presentado en los ejemplos anteriores se puede generalizar 
  de la siguiente forma. 
  
  Sea $f:R^n\rightarrow R^m$ una función diferenciable. Se denotarán las $n$ 
  variables reales de entrada como
  \begin{displaymath}
	 w_{i-n}=x_i\qquad \textrm{y}\qquad \dot{w}_{i-n}=\dot{x}_i,\ \ 
	 \textrm{con } i=1\ldots n \textrm{\ \ y \ } \dot{x}_i=1.
  \end{displaymath}
  Las variables $\dot{x}_i$ se inicializan con el valor de las
  derivadas parciales de cada variable $x_i$ con respecto a la variable original.
  Si se desea calcular la derivada parcial de $f$ respecto a $x_i$, entonces 
  $\dot{x}_i$ deber'ia ser $1$ y $\dot{x}_j$ deber'ia ser $0$ para toda $j$ 
  diferente de $i$.
  
  Como $f$ está compuesta por funciones elementales, es conveniente denotarlas de alguna forma. 
  A la $j$-ésima función elemental se le denotará por $\phi_j$. 
  
  Para descomponer a $f$ en operaciones 
  elementales se denotarán nuevas variables $w_i$ y $\dot{w}_i$ como
  
  \begin{equation}
   w_i=\phi_i(w_j)\qquad\textrm{con\ }j\prec i, \label{eq:notaciondependencia}
  \end{equation}
  
  \begin{displaymath}
	 \dot{w}_i=\sum_{j\prec i}\frac{\partial}{\partial w_j}\phi_i(w_j)\dot{w}_j,
  \end{displaymath}

  donde en este caso $i=1\ldots l$ y $l$ es el número de operaciones elementales que componen a $f$. 
  La simbología $\phi_i(w_j)$ con $j\prec i$ significa que $\phi_i$ depende directamente de $w_j$, 
  es decir, que para evaluar $\phi_i$ se usa explícitamente $w_j$. También es usual denotar 
  (\ref{eq:notaciondependencia}) como
  
  \begin{displaymath}
	 w_i=\phi_i(w_j)_{j\prec i}.
  \end{displaymath}

  Con las notaciones anteriores se puede expresar el procedimiento
  general de la siguiente forma:

	\begin{displaymath}
	 \begin{tabular}{|l l|}
	  \hline
	  $w_i=x_i$ & \\
	  & $i=1\ldots n$\\
	  $\dot{w}_{i-n}=\dot{x}_i$ & \\
	  \hline
	  $w_i=\phi_i(w_j)_{j\prec i}$ & \\
	  & $i=1\ldots l$\\
	  $\dot{w}_i=\sum_{j\prec i}\frac{\partial}{\partial w_j}\phi_i(w_j)\dot{w}_j$ & \\
	  \hline
	  $y_{m-i}=w_{l-i}$ & \\
	  & $i=m-1\ldots 0$\\
	  $\dot{y}_{m-i}=\dot{w}_{l-i}$ & \\
	  \hline
	 \end{tabular}
  \end{displaymath}
  
  Como se puede apreciar en la tabla anterior, el procedimiento general del modo forward 
  est'a estructurado en tres fases: primero se inicializan las variables, despu'es se 
  ejecutan las operaciones y se calculan las derivadas, y finalmente se devuelven los 
  valores y la derivada calculada.
  
  En esta secci'on se ha presentado la AD desde un punto de vista te'orico. En la siguiente
  secci'on se muestra una posible v'ia para implementar estas ideas computacionalmente.

 \subsection{Implementación} \label{sec:implementacion}
 
 Existen dos estrategias para lograr AD: transformación del 
 código fuente y sobrecarga de operadores \cite{griewank}.

 Para utilizar la metodolog'ia propuesta en este trabajo resulta m'as conveniente 
 utilizar la sobrecarga de operadores, por lo que en esta secci'on se presentan sus
 elementos fundamentales. El lector interesado en la transformaci'on de c'odigo puede
 consultar \cite{griewank1}. 

 Para implementar la AD utilizando sobrecarga de operadores se debe crear una nueva clase, 
 la cual se llamar'a en este trabajo Adouble siguiendo la notaci'on de \cite{griewank}.
  
 En esta clase se deben definir dos campos de números reales, uno, que usualmente recibe 
 el nombre de valor, para almacenar el resultado de la operaci'on que este Adouble
 representa, y otro, que usualmente recibe el nombre de derivada, para almacenar la
 derivada de esa operaci'on. Estos campos valor y derivada son la representaci'on 
 computacional de las variables $w_i$ y $\dot{w}_i$. 
  
 Una vez definida esta nueva clase, se sobrecargan
 los operadores aritm'eticos y las funciones elementales, para que, al mismo tiempo que
 se calculan los resultados de las operaciones, tambi'en se pueda calcular el valor de 
 las derivadas.
 
 La siguiente tabla muestra los valores de los campos valor y derivada
 de cada uno de los Adoubles que intervienen en la evaluaci'on de la funci'on
 $f(x) = x^2\cdot{}cos(x)$. N'otese que cuando se realizan todas las operaciones, en el
 campo derivada del Adouble $y$ se obtiene el valor de la derivada de la funci'on
 en el punto en que fue evaluada.
 
 \begin{displaymath}
  \begin{tabular}{|l|l|l|}
   \hline
   Adouble $w_1$ & $w_1$.valor $= \pi$ & $w_1$.derivada $= 1$ \\
   \hline
   Adouble $w_2 = w_1^2$ & $w_2$.valor $=\pi^2$ & $w_2$.derivada $= 2\pi$ \\
   Adouble $w_3 = \cos(w_1)$ & $w_3$.valor $=-1$ & $w_3$.derivada $=0$\\
   Adouble $w_4 = w_2*w_3$ & $w_4$.valor $=-\pi^2$ & $w_4$.derivada $=-2\pi$\\
   \hline
   Adouble\ \ \ $y = w_4$ &\ \ $y$.valor\ $=-\pi^2$ &\ \  $y$.derivada\ $=-2\pi$\\
   \hline
  \end{tabular}
 \end{displaymath} 
  
 Esta vía tiene la ventaja de que es f'acil de implementar y que para utilizarlo solo hay
 que modificar ligeramente el c'odigo fuente de la funci'on que se desea derivar \cite{griewank1}.

 En la Figura \ref{cod1} se muestra la implementaci'on de un programa en C\# que calcula el valor de 
 $f(x) = x^2\cdot\cos(x)$ en $x = 2$; y en la Figura \ref{cod2} las modificaciones necesarias para calcular la derivada en ese punto.
 
 \begin{figure}[htb]
  \begin{verbatim}
class Example
{
  static void Main()	
  {
    double x = 2; //Inicialización de x
                  //para evaluar f en 2
    double y = x*x + cos(x); //Cálculo de f
    Console.WriteLine("El valor de f en x = 2 es: " + y);
  }
}
  \end{verbatim}
	\caption{Código de $f(x) = x^2\cdot\cos(x)$}\label{cod1}
 \end{figure}
 
 \begin{figure}[htb]
  \begin{verbatim}
class Example
{
  static void Main()	
  {
    Adouble x = new Adouble(); \\Inicialización de x como 
                               \\Adouble
    x.valor = 2; \\ Se quiere el valor y la derivada
                 \\ en x = 2
    x.derivada = 1; \\La derivada de x con respecto a x es 1
    Adouble y = x*x + cos(x);
    Console.WriteLine("El valor de f en x = 2 es: " + w4.valor);
    Console.WriteLine("La derivada de f en x = 2 es: " + w4.derivada);
  }
}
  \end{verbatim}
	\caption{Código modificado de $f(x) = x^2\cdot\cos(x)$}\label{cod2}
 \end{figure}
 
 Esta idea se puede modificar f'acilmente para calcular derivadas parciales 
 de funciones de m'as de una variable. El lector interesado en este tema puede consultar
 \cite{griewank} y \cite{adolc}.
  
\section{C'alculo de Derivadas Anidadas} \label{cha:metodologia}

 En esta secci'on se presenta una metodolog'ia para calcular derivadas anidadas
 utilizando cualquier biblioteca de Diferenciaci'on Autom'atica mediante sobrecarga 
 de operadores. Esta metodolog'ia se presenta para funciones de una variable real
 para a facilitar su exposici'on, pero su extensi'on a funciones de varias variables
 es posible de la misma forma que las t'ecnicas de diferenciaci'on autom'atica se 
 extienden a funciones varias variables \cite{griewank}.

 La metodolog'ia que se propone es la siguiente: 
 
 \begin{enumerate}
  \item Partir de una biblioteca de Diferenciaci'on Autom'atica en la que exista
        un tipo de dato Adouble.
  \item Definir un nuevo tipo de dato llamado SuperAdouble. Esta nueva estructura tendr'a 
        dos campos: valor y derivada, y ambos campos ser'an del tipo de dato Adouble
        definido en la biblioteca del paso 1. El hecho de que 
        estos campos sean de tipo Adouble permitir'a calcular las derivadas anidadas.
  \item Definir funciones para construir SuperAdoubles a partir de Adoubles y para obtener
        los campos valor y derivada de variable de tipo SuperAdouble. Estas funciones se presentan en la 
        Secci'on \ref{paso2}.
  \item Modificar el c'odigo fuente de la funci'on anidada para calcular las derivadas anidadas.
 \end{enumerate}      
 
 En las siguientes secciones se detallan cada uno estos pasos.
 
 \subsection{La clase SuperAdouble} \label{paso1}
 
 El segundo paso propone crear una nueva clase llamada SuperAdouble. Este nuevo tipo de dato tiene un
 campo valor y un campo derivada, ambos de tipo Adouble. Al igual que la clase Adouble, los operadores
 aritm'eticos y las funciones elementales se sobrecargan para que cuando operen con valores de este tipo
 permitan calcular los valores y las derivadas anidadas que se deseen. 
          
 El campo valor de los SuperAdouble es de tipo Adouble, por lo que este contiene
 a un campo valor y un campo derivada, ambos de tipo double. Es decir, 
 para una variable de tipo SuperAdouble tiene sentido hablar de los campos 
 valor.valor y valor.derivada. Por el mismo
 motivo, en una variable de tipo SuperAdouble existen los campos derivada.valor y derivada.derivada.
 
 Estos cuatro campos: valor.valor, valor.derivada, derivada.valor y derivada.derivada 
 se relacionan con las derivadas anidadas a trav'es del siguiente resultado, que es 
 el principal aporte de este trabajo.
 
 \begin{teorema} \label{teo:teorema}
  En un objeto de tipo SuperAdouble se cumple que:
  \begin{itemize}
 	 \item valor.valor es el valor de la funci'on anidada. 
	 \item valor.derivada es la derivada original.
	 \item derivada.valor contiene la derivada anidada.
	 \item derivada.derivada contiene la derivada compuesta.
  \end{itemize}
 \end{teorema}
 \begin{proof}
  La demostraci'on se encuentra en \cite{tesis}.
 \end{proof}
 
 Una vez que est'e definida esta nueva clase, es necesario definir funciones
 para crear instancias de esta clase y obtener las derivadas y valores deseados.
 Estas funciones se muestran en la siguiente secci'on.
              
 \subsection{Relaci'on entre Adoubles y SuperAdoubles.} \label{paso2}
              
 Para calcular derivadas anidadas son necesarias tres funciones que relacionen
 variables de tipo Adouble y variables de tipo SuperAdouble. Estas funciones 
 son \emph{push}, \emph{popV} y \emph{popD}. A continuaci'on se presentan 
 cada una de ellas y su relaci'on con el c'alculo de derivadas anidadas.
 
 La función \emph{push} recibe como par'ametro una variable $x$ de tipo Adouble y devuelve 
 una variable $X$ de tipo SuperAdouble. El campo valor de la
 variable $X$ ser'ia el Adouble $x$, y el campo derivada ser'ia un Adouble con valor
 1 y derivada 0. 
 
 \begin{ejemplo}
  Si se tiene un Adouble $x = (4, 10)$, donde la primera componente del vector es el campo valor del
  Adouble, y la segunda, el campo derivada, entonces al aplicar push
  se obtiene un SuperAdouble $X$ con los siguientes campos:
  \begin{itemize}
 	 \item $X.$valor.valor = 4.
	 \item $X.$valor.derivada = 10.
	 \item $X.$derivada.valor = 1.
	 \item $X.$derivada.derivada = 0.
  \end{itemize}
 \end{ejemplo}
 
 Por otra parte, las funciones \emph{popV} y \emph{popD} reciben como argumento un 
 SuperAdouble y devuelven un Adouble. La funci'on \emph{popV} devuelve el campo valor
 del SuperAdouble y \emph{popD} devuelve su campo derivada. 
 
 \begin{ejemplo}
  Si se aplica la funci'on popV al SuperAdouble $X$ definido en el ejemplo
  anterior se obtiene un Adouble $x$ con los siguientes campos:
  \begin{itemize}
   \item $x.$valor = $4$.
	 \item $x.$derivada = $10$.
  \end{itemize} 
 
  Por otra parte, si se aplica la funci'on popD al
  mismo SuperAdouble $X$ se obtiene:
  \begin{itemize}
	 \item $x.$valor = $1$.
	 \item $x.$derivada = $0$.
  \end{itemize} 
 \end{ejemplo}
 
 Para calcular derivadas anidadas, estas funciones deben usarse para modificar el 
 c'odigo fuente de la funci'on anidada como se muestra en la siguiente
 secci'on.
  
 \subsection{Modificaci'on del c'odigo de la funci'on anidada}
 
 En esta secci'on se presentan las modificaciones que son necesarias realizar
 en el c'odigo fuente de la funci'on anidada para utilizar los SuperAdouble, las
 funciones \emph{push, popD} y \emph{popV} y obtener los valores de las derivadas anidadas.
  
 Siguiendo con el Ejemplo \ref{ej:anidaad},
 \begin{displaymath}
   f(x)= x^2+\dot{g}(x^3), \qquad g(x)=e^{x^2}, 
 \end{displaymath}	
 la Figura \ref{fig:modif} muestra el c'odigo que permite calcular 
 el valor de la funci'on y su derivada en el punto $x=3$ usando los SuperAdoubles y las funciones 
 presentadas en la secci'on anterior.
 
 \begin{figure}[htb]
  \begin{verbatim}
class Example
{
  static void Main()	
  {
    Adouble x = new Adouble();
    x.valor = 3;
    x.derivada = 1;
    Adouble w1 = x*x;             \\ w1 = x^2
    Adouble w2 = w1*x;            \\ w2 = x^3
    SuperAdouble Y1 = push(w2);   \\ crear el SuperAdouble
    SuperAdouble Y2 = exp(Y1*Y1); \\ g(x) = exp(x^2)
    SuperAdouble W3 = exp(W2);
    Adouble w3 = popD(Y2);        \\ obtener la derivada anidada
    Adouble w4 = w1 + w3;         \\ w4 = x^2 + gdot(x^3)
    Console.WriteLine("El valor de f en x = 3 es: " + 
                       w5.valor);
    Console.WriteLine("La derivada de f en x = 3 es: " + 
                       w5.derivada);
  }
}  
  \end{verbatim}
  \caption{Aplicaci'on de los SuperAdoubles en la AD.} \label{fig:modif}
 \end{figure}
 
 En el c'odigo de la Figura \ref{fig:modif} se empieza trabajando con
 números de tipo Adouble para calcular el valor y la derivada de las
 operaciones que no pertenezcan a la funci'on anidada. La l'inea
 \begin{verbatim}
  SuperAdouble Y1 = push(w2);
 \end{verbatim}
 \vspace*{-0.5cm}
 crea un objeto SuperAdouble a partir de la variable anidada. Despu'es
 de la creaci'on de W1 se efect'uan las operaciones de la funci'on anidada utilizando
 SuperAdoubles. Con la l'inea
 \begin{verbatim}
  Adouble w3 = popD(Y2);
 \end{verbatim}
 \vspace*{-0.5cm}
 se obtiene un Adouble que en su campo valor tiene la
 derivada anidada, y en su campo derivada tiene la derivada
 compuesta. Finalmente, en la l'inea
 \begin{verbatim}
  Adouble w4 = w1 + w3;
 \end{verbatim}
 \vspace*{-0.5cm} se termina el c'alculo de la funcion original.
 
 Como todas las operaciones se han realizado con Adoubles, en el campo
 derivada del Adouble w4 se tiene la derivada de la funci'on original, que era
 el objetivo que se persegu'ia.

\section{Conclusiones}

 Con la metodología propuesta en este trabajo es posible utilizar la
 Diferenciación Automática para calcular derivadas anidadas. Esta
 metodología es sencilla de implementar gracias a que se puede reusar
 otras librerías de AD que soporten sobrecarga de operadores.

 Como recomendaciones y trabajo futuro se propone implementar el modo
 hacia atrás de la Diferenciación Automática con los números de tipo
 SuperAdouble, lo cual sería útil en los casos que la función anidada
 con dominio en $\mathbb{R}^n$ e imagen en $\mathbb{R}$, y generalizar
 los resultados obtenidos para el caso en que haya más de un nivel de
 anidación.


\end{document}